\documentclass[showpacs,preprintnumbers,amsmath,aps,amssymb]{revtex4}
\usepackage{graphicx}
\usepackage{dcolumn}
\usepackage{bm}
\begin{document}

\baselineskip 17pt

\title{Anisotropic cosmology in S\'aez-Ballester theory: \\
classical and quantum solutions}
\author{J. Socorro$^1$ }
\email{socorro@fisica.ugto.mx}
\author{M. Sabido$^1$ }
\email{msabido@fisica.ugto.mx}
\author{M.A. S\'anchez G.$^2$}
\author{ M.G. Fr\'{\i}as Palos$^2$}
\affiliation{$^1$Departamento de F\'{\i}sica, DCeI, Universidad de 
Guanajuato-Campus Le\'on\\
A.P. E-143, C.P. 37150,  Guanajuato, M\'exico\\
$^2$ Facultad de Ciencias de la Universidad Aut\'onoma del Estado de 
M\'exico,\\
Instituto Literario  No. 100, Toluca, C.P. 50000, Edo de Mex, M\'exico}%

\begin{abstract}
We use the  S\'aez-Ballester theory on  anisotropic Bianchi I cosmological model, 
with barotropic fluid and cosmological constant. We obtain the classical solution by using the Hamilton-Jacobi approach. 
Also the quantum regime is constructed and exact solutions to the Wheeler-DeWitt equation are found. \\


Usamos la teor\'{\i}a de S\'aez-Ballester en el modelo anisotr\'opico Bianchi I con un fluido barotr\'opico y constante
cosmol\'ogica. Obtenemos las soluciones cl\'asicas usando el enfoque de Hamilton-Jacobi. El regimen cu\'antico tambi\'en
es construido y soluciones exactas a la ecuaci\'on de Wheeler-DeWitt son encontradas.\\

Descriptores: Soluciones cl\'asicas y cu\'anticas exactas; cosmolog\'{\i}a.
  \end{abstract}

\maketitle
\section{Introduction}
Saez and Ballester \cite{s-b} formulated a scalar-tensor theory of gravitation
in which the metric is coupled with a dimensionless scalar field. In this direction, many works in the classical regime have been done
\cite{singh,shri,mohanty,singh-shri}, yet a study of the anisotropy behaviour trough the form  introduced in 
the line element has been conected. In  this theory the strenght of the coupling between gravity and the scalar field
is determined by an arbitrary coupling function $\omega$. In spite
the dimensionless character of the scalar field, an antigravity regime appears,   this  suggests a possible way to solve
the missing matter problem in non-flat FRW cosmologies. However, Armendariz-Picon et al called this scenario as {\it K-essence} 
\cite{armendariz}, which is characterized by a scalar field with a non-canonical kinetic energy. Usually K-essence models
are restricted to the lagrangian density of the form
\begin{equation}
\rm S=\int d^4x \, \sqrt{-g}\, f(\phi) \, \left(\nabla \phi \right)^2,
\end{equation}
one of the motivations to consider this type of lagrangian originates from string theory \cite{string}. 
For more details for K-essence applied
to dark energy,  you can  see \cite{copeland} and reference therein. 

 On another front, the quantization program of this theory has not ben constructed, 
because should be dark
to build the ADM formalism. Thus, we transform this theory to conventional one, where the dimensionless scalar field is obtained from
energy-momentum tensor as a exotic matter, and in this sense, we can use this structure for the quantization program, where the ADM formalism is well
known for different classes of matter \cite{ryan1}

In this work, we use this formulation  to obtain classical and quantum exact solutions to anisotropic Bianchi type I cosmological model, 
including a cosmological constant $\Lambda$. 
 The first step is to write S\'aez-Ballester formalism in the usual manner, that is, we calculate the
corresponding energy-momentum tensor to the scalar field and give the equivalent lagrangian density. Next, we proceed to obtain
the corresponding canonical lagrangian ${\cal L}_{can}$ to  Bianchi type I through the lagrange transformation, 
we calculate the classical hamiltonian 
${\cal H}$, from which we find the Wheeler-DeWitt (WDW) equation of the corresponding cosmological model under study. We employ in this work
the Misner parametrization due that in natural way appear the anisotropy parameters to the scale factors, and we can to analyze its
behaviour in  easy way.

The more simple generalization to lagrangian density for the S\'aez-Ballester theory  \cite{s-b} with cosmological term, is  
\begin{equation}
\rm {\cal L}_{geo}=\left( R- 2\Lambda - F(\phi) \phi_{,\gamma} \phi^{,\gamma}\right), \label{lagrangian}
\end{equation}
where $\phi^{,\gamma}=g^{\gamma \alpha} \phi_{,\alpha}$, R the scalar curvature, $F(\phi)$ a dimensionless functional scalar field. 
In classical field theory with scalar field, this formalism corresponds to null potencial in the field $\phi$, but the
kinetic term is exotic by the factor $F(\phi)$. 

From the lagrangian (\ref{lagrangian}) we can build the complete action
\begin{equation}
\rm I=\int_{\Sigma} \sqrt{-g}({\cal L}_{geo}+ {\cal L}_{mat})d^4x,\label{action}
\end{equation}
where ${\cal L}_{mat}$ is the matter lagrangian,  g is the determinant of metric tensor.
The field equations for this theory are
\begin{eqnarray}
\rm G_{\alpha \beta}+g_{\alpha\beta}\Lambda-F(\phi) \left( \phi_{,\alpha}\phi_{,\beta} - \frac{1}{2} g_{\alpha \beta} \phi_{,\gamma} \phi^{,\gamma} \right)
&=& \rm 8\pi G T_{\alpha \beta}, \nonumber \\
\rm 2F(\phi) \phi^{,\alpha}_{\,\,;\alpha} + \frac{dF}{d\phi}\phi_{,\gamma} \phi^{,\gamma}&=&0, \label{fe}
\end{eqnarray}
 where G is the gravitational constant and as usual the semicolon means a covariant derivative.

These set of equations (\ref{fe}) can be obtained using the equivalent lagrangian as a matter and energy-momentum tensor for this field $\phi$,
\begin{eqnarray}
\rm {\cal L}_{\phi}&=&\rm F(\phi) g^{\alpha \beta}\phi_{,\alpha} \phi_{,\beta},\nonumber\\
\rm T_{\alpha\beta}(\phi)&=&\rm F(\phi) \left(\phi_{,\alpha}\phi_{,\beta}- \frac{1}{2} g_{\alpha \beta} \phi_{,\gamma} \phi^{,\gamma}\right),
\end{eqnarray}

in this way, we write the action (\ref{action}) in the usual form
\begin{equation}
\rm I=\int_{\Sigma} \sqrt{-g}\left( R- 2\Lambda +{\cal L}_{mat}+{\cal L}_\phi \right)d^4x,\label{action1}
\end{equation}
and consequently, the classical equivalence between the two theories. We can infer that this correspondence also is satisfied in the quantum 
regimen, because only is modified the hamiltonian constraint \cite{ryan1}.

This work is arrangede as follow. In section II we construct the hamiltonian density for the cosmological model. In
section III  the classical  solutions using the Hamilton-Jacobi formalism are found. Here, we have used a barotropic perfect fluid as a matter content and a cosmological constant, 
obtaining the solutions for differents epoch of the evolution for this cosmological model. In Section IV the
cuantization scheme, obtaining the corresponding Wheeler-DeWitt equation and its solutions for different values for the $\gamma$ parameter. Finally, 
the section V is devoted to discussion.

\section{The hamiltonian density}
The line element for the cosmological Bianchi type I has the form
\begin{equation}
\rm ds^{2} =-N^2dt^{2}+ e^{ 2\Omega+ 2\beta _{+}+2\sqrt{3}\beta _{-}} \left( dx^{1}\right) ^{2}+e^{2\Omega
+ 2\beta _{+}-2\sqrt{3}\beta _{-}} \left(
dx^{2}\right) ^{2} +e^{2\Omega -4\beta _{+}} \left(
dx^{3}\right) ^{2}
\end{equation}
where N is the lapse function, $\beta_\pm$ are the corresponding anisotropic parameter in the scale factors, $\Omega$ play the role
as the scale factor like to flat Friedmann-Robertson-Walker cosmological model $(e^\Omega\equiv A)$. The total volume for all diagonal
Bianchi cosmological models is given by the expression $\rm V=e^{3\Omega(t)},$that will appear in the solutions for all parameter in this theory.

Then,  the corresponding lagrangian density in this theory is
\begin{equation}
{\cal L}=\rm \frac{6\dot{ \Omega^2} e^{3\Omega}}{N}-6\frac{\dot \beta _+^2 e^{ 3\Omega} }{N}-6\frac{\dot\beta_-^2 e^{3\Omega}}{N}+
\frac{F(\phi)}{N} \dot\phi^2 e^{ 3\Omega} 
+\left( 16N \pi G\rho  -2N\Lambda \right) e^{3\Omega}. \nonumber
\end{equation}
wich can be rewritten in the canonical form,\begin{equation}
{\cal L} _{can}= \rm \Pi _{\Omega }\overset{\cdot }{\Omega }+\Pi _{\beta
_{+}}\overset{\cdot }{\beta _{+}}+\Pi _{\beta _{-}}\overset{\cdot }{\beta
_{-}}+\Pi _{\phi }\overset{\cdot }{\phi }-N{\cal H}, \label{canon}
\end{equation}
with ${\cal H}$ as the hamiltonian density, and the momentas are defined in the usual way 
$\Pi _{q^i}=\rm \frac{\partial {\cal L} }{\partial \dot q^i}$, where $\rm q^i=(\Omega, \beta_+, \beta_-, \phi)$ are
the field coordinates for this system,
\begin{eqnarray}
\rm \Pi _{\Omega } &=&\rm \frac{\partial {\cal L} }{\partial \dot \Omega }=12\frac{\dot \Omega e^{3\Omega }}{N}, 
\qquad \to \qquad \frac{\Pi _{\Omega }}{2} =6\frac{\dot \Omega e^{ 3\Omega} }{N} \quad \to \quad 
\dot \Omega  =Ne^{-3\Omega }\frac{\Pi _{\Omega }}{12},  \label{omega}\\
\rm \Pi _{\beta _{+}} &=&\rm \frac{\partial {\cal L} }{\partial \dot \beta _+}=-12\frac{\dot\beta_+ e^{3\Omega} }{N},\quad \to 
\quad \frac{\Pi _{\beta _+}}{2} =-6\frac{\dot \beta_+ e^{3\Omega} }{N} \quad \to \quad 
\dot\beta_+=-Ne^{-3\Omega}\frac{\Pi_{\beta_+}}{12}, \label{+}\\
\rm \Pi _{\beta _{-}} &=&\rm \frac{\partial {\cal L} }{\partial \dot \beta _-}=-12\frac{\dot\beta_- e^{3\Omega} }{N},\quad \to 
\quad \frac{\Pi _{\beta _-}}{2} =-6\frac{\dot \beta_- e^{3\Omega} }{N} \quad \to \quad 
\dot\beta_-=-Ne^{-3\Omega}\frac{\Pi_{\beta_-}}{12},\label{-}\\
\rm \Pi _{\phi } &=&\rm \frac{\partial {\cal L} }{\partial \dot \phi }=2\frac{F(\phi) \dot\phi e^{3\Omega} }{N}, \qquad \to 
\quad \frac{\Pi_\phi}{2} =\frac{F(\phi) \dot\phi e^{3\Omega }}{N} \quad \to \quad 
\dot \phi  =\frac{N\Pi _{\phi }}{2F(\phi) e^{3\Omega} }. \label{momentos}
\end{eqnarray}

The matter is introduced as a barotropic perfect fluid $P=\gamma \rho$ with $\gamma$ a constant between $-1<\gamma<1$,
 who energy-momentum tensor is given by the following expression
$\rm T_{\mu\nu}= (\rho + P)U_\mu U_\nu - g_{\mu\nu} P$
where $\rm U_\mu$ is the  four-velocity, $\rho$ is the energy density and P is the thermodynamic pressure in the fluid, respectively.
Using the covariant estructure in this tensor, we obtain the differential equation
$\rm 3\dot \Omega \rho +3\dot \Omega P+\dot \rho=0$
and the solution
\begin{equation}
\rm \rho =\mu_\gamma e^{-3\Omega \left( 1+\gamma \right)}. \nonumber
\end{equation}
with $\mu_\gamma$ a constant for the corresponding scenario.

So, we obtain the following expresion for the canonical lagrangian
\begin{equation}
{\cal L} _{can}=\rm \Pi _{q^{i}}\dot q^i-\frac{N e^{-3\Omega} }{24}\left[ \Pi _{\Omega }^{2}-\Pi _{\beta _{+}}^{2}-\Pi _{\beta
_{-}}^{2}+\frac{6\Pi _{\phi }^{2}}{F(\phi) }-384\pi G\mu_\gamma 
e^{3\Omega \left( 1-\gamma \right)} +48\Lambda e^{ 6\Omega} \right], \nonumber
\end{equation}
and
\begin{equation}
{\cal H}=\rm \frac{e^{-3\Omega} }{24}\left( \Pi _{\Omega }^{2}-\Pi _{\beta
_{+}}^{2}-\Pi _{\beta _{-}}^{2}+\frac{6\Pi _{\phi }^{2}}{F(\phi) }%
-384\pi G\mu_\gamma e^{3\Omega \left( 1-\gamma \right)} +48\Lambda e^{ 6\Omega} \right),
\end{equation}
and using the lagrange equation for the field N,  we get the Hamiltonian constraint
\begin{equation}
{\cal H}=\rm \Pi _{\Omega }^{2}-\Pi _{\beta _{+}}^{2}-\Pi _{\beta _{-}}^{2}+\frac{6\Pi
_{\phi }^{2}}{F(\phi) }-384\pi G\mu_\gamma e^{3\Omega \left( 1-\gamma\right)} +48\Lambda e^{6\Omega} =0.
\end{equation}

\section{Classical regimen: Hamilton-Jacobi approach}
Employing the  Hamilton-Jacobi formulation, where the momentas
are  $\rm \Pi_q=\frac{\partial S_q}{\partial q}$, where $\rm S_q$ is the superpotential function, the 
hamiltonian takes the following form
\begin{equation}\rm
\left( \frac{\partial S_{\Omega }}{\partial \Omega }\right) ^{2}-\left( 
\frac{\partial S_{\beta _{+}}}{\partial \beta _{+}}\right) ^{2}-\left( \frac{%
\partial S_{\beta _{-}}}{\partial \beta _{-}}\right) ^{2}+\frac{6}{F(\phi) }\left( \frac{\partial S_{\phi }}{\partial \phi }\right) ^{2}-384\pi
G\mu_\gamma e^{3\Omega \left( 1-\gamma \right)} +48\Lambda e^{6\Omega} =0,
\end{equation}
solving for the variable $\Omega$ we have
\begin{equation*}
\rm \left( \frac{\partial S_{\Omega }}{\partial \Omega }\right) ^{2}-384\pi G\mu_\gamma
e^{3\Omega \left( 1-\gamma \right)} +48\Lambda e^{6\Omega} =\left( \frac{%
\partial S_{\beta _{+}}}{\partial \beta _{+}}\right) ^{2}+\left( \frac{%
\partial S_{\beta _{-}}}{\partial \beta _{-}}\right) ^{2}-\frac{6}{F(\phi) }\left( \frac{\partial S_{\phi }}{\partial \phi }\right) ^{2}=\xi ^{2}
\end{equation*}
using (\ref{omega}), we obtain the following integral equation depending on time parameter
$\rm Ndt=d\tau$, we have
\begin{eqnarray}
\rm \frac{dS_\Omega }{d\Omega } &=&\rm \sqrt{\xi^{2}+384\pi G\mu_\gamma 
e^{3\Omega\left( 1-\gamma \right)} -48\Lambda e^{6\Omega} }=12e^{3\Omega} \frac{d\Omega }{d\tau } \\
\rm \Delta \tau &=&\rm \int \frac{12 }{\sqrt{\xi ^{2}e^{-6\Omega}+384\pi G\mu_\gamma
e^{- 3\Omega \left( 1+\gamma \right)} -48\Lambda}} d\Omega
\end{eqnarray}
where $\xi $ is a separation constant. This equation do not have a general solution, but we can solve for the particular values
of the $\gamma$ parameter, which we present in the Table I.

\bigskip\bigskip
\begin{center}
\begin{tabular}{|l|c|}\hline
Case  & $\rm \Omega(\tau)$ \\ \hline
Inflation, $\gamma=-1$ & 
$\rm \frac{1}{3}\, Ln \,
\left[ \frac{1}{4\sqrt{b_{-1}}}\left( e^{\frac{\sqrt{b_{-1}}}{4}\Delta \tau}-4\xi^2e^{-\frac{\sqrt{b_{-1}}}{4}\Delta \tau}\right)\right]$ \\ 
$\rm b_{-1}=384\pi G \mu_{-1}-48\Lambda>0$ & \\\hline
Dust, $\gamma=0$ & $\rm \frac{1}{3}
Ln\,\left[\frac{\left(e^{\sqrt{-3\Lambda}\Delta \tau}-\frac{b_0}{4\sqrt{-3\Lambda}} \right)^2
-4\xi^2}{16\sqrt{-3 \Lambda}e^{\sqrt{-3\Lambda}\Delta \tau}} \right] $\\
$\rm b_0=384\pi G \mu_0$, \, $\Lambda<0$ & \\\hline
Stiff matter, $\gamma=1$ & $\rm \frac{1}{3}\, 
Ln \,\left[ \frac{1}{16\sqrt{-3\Lambda}}\left( e^{\sqrt{-3\Lambda}\Delta \tau}-4b_1 e^{-\sqrt{-3\Lambda}\Delta \tau}\right)\right]$\\
$\rm b_1=\xi^2+384\pi G \mu_1$, $\Lambda <0$ & \\\hline
\end{tabular} \\
\emph{ Table I. Solutions for $\Omega$ in the scenarios
$\gamma=-1,0,1$}
\end{center}
The corresponding solutions for the anisotropic functions and the field $\phi$ appear in the Table II. For the field $\phi $
we consider
\begin{eqnarray*}
\rm \frac{6}{F(\phi) }\left( \frac{\partial S_{\phi }}{\partial \phi }%
\right) ^{2} &=&\rm -\xi ^{2}+\kappa_+ ^{2}+\kappa_-^{2}=\theta^{2} \\
\frac{dS_{\phi }}{d\phi } &=&\sqrt{\frac{F(\phi) }{6}}\theta \equiv -2F(\phi)\frac{d\phi}{d\tau} e^{3\Omega}  
\end{eqnarray*}
with  $\rm \theta^2=-\xi ^{2}+\kappa_+ ^{2}+\kappa_-^{2} $, with the cuadrature solutions
\begin{equation}
\rm \int \sqrt{F(\phi)}\, d\phi= \frac{\theta}{2\sqrt{6}}\int e^{-3\Omega}(\tau)\, d\tau, \label{phi-new}
\end{equation}

Now considering the original S\'aez-Ballester theory, $\rm F(\phi)=\omega \phi^m$,
the corresponding solution for all m are

\bigskip\bigskip
\begin{center}
\begin{tabular}{|l|l|l|}\hline
Case  & $\beta_\pm(\tau)$ &$\phi(\tau)$\\ \hline
$\gamma=-1$, $\omega>0$ & 
$\rm  \mp \frac{2 \kappa_\pm }{3\xi}\, tanh^{-1}\, \left(\frac{e^{\frac{\sqrt{b_{-1}}}{4}\Delta \tau}}{2\xi}  \right)$&
$\rm  \begin{tabular}{lr}
 $\rm \left[ \mp \frac{2\theta (m+2)}{\xi  \sqrt{6\omega}}\, 
 tanh^{-1}\, \left(\frac{Exp\left(\frac{\sqrt{b_{-1}}}{4}\Delta \tau\right)}{2\xi}  \right)\right]^{\frac{2}{m+2}},$&  $m\not= -2 $\\
 $ \rm Exp\left[\mp \frac{4\theta }{\xi \sqrt{6\omega }}  tanh^{-1}\, \left(\frac{Exp\left(\frac{\sqrt{b_{-1}}}{4}\Delta \tau\right)}{2\xi}  
 \right)\right],$&  m=-2 \\
 \end{tabular}   $ \\ 
 $k_+=-\sqrt{3}k_-$ & & \\\hline
$\gamma=0$ & $\rm  \pm \frac{2\kappa_\pm}{3\xi}\, tanh^{-1}\, 
\left(\frac{-b_0 +4\sqrt{-3\Lambda} e^{\sqrt{-3\Lambda}\Delta \tau}}{8\sqrt{-3\Lambda}\xi} \right) $&
$\rm  \begin{tabular}{lr}
 $\rm \left[\pm \frac{2\theta (m+2)}{\xi \sqrt{6\omega}} tanh^{-1}\, \left(\frac{-b_0 +4\sqrt{-3\Lambda} 
 e^{\sqrt{-3\Lambda}\Delta \tau}}{8\sqrt{-3\Lambda}\xi}  \right)\right]^{\frac{2}{m+2}}\, ,$&  $m\not= -2$\\
 $\rm Exp\left[\pm \frac{4\theta}{\xi \sqrt{6\omega}}\,tanh^{-1}\, 
 \left(\frac{-b_0 +4\sqrt{-3\Lambda} e^{\sqrt{-3\Lambda}\Delta \tau}}{8\sqrt{-3\Lambda}\xi}  \right)\right], $ &  m=-2,\\
 \end{tabular}$  \\
$\omega>0$   & &\\\hline
$\gamma=1$, $\omega>0$ & $\rm  \mp \frac{2\kappa_\pm}{3\sqrt{b_1}} \, 
tanh^{-1}\left( \frac{e^{ \sqrt{-3\Lambda}\tau}}{2\sqrt{b_1}}\right)$ &
$\rm  \begin{tabular}{lr}
 $\rm \left[ \mp \frac{2\theta (m+2)}{  \sqrt{6b_1 \omega}}\, 
 tanh^{-1}\, \left(\frac{e^{\sqrt{-3\Lambda}\Delta \tau}}{2\sqrt{b_1}}  \right)\right]^{\frac{2}{m+2}},$&  $m\not= -2 $\\
 $ \rm Exp\left[\mp \frac{4\theta }{ \sqrt{6b_1 \omega }}  tanh^{-1}\, \left(\frac{e^{\sqrt{-3\Lambda}\Delta \tau}}{2\sqrt{b_1}}  
 \right)\right],$&  m=-2 \\
  \end{tabular}   $ \\
  $k_+=-\sqrt{3} k_-$ & & \\\hline
\end{tabular} \\
\emph{ Table II. Solutions for the anisotropic variables $\beta_\pm$ and field $\phi$ in the scenarios
$\gamma=-1,0,1$}
\end{center}

These set of solutions satisfy the Einstein field equation (\ref{fe}), which were checked using REDUCE package.
It is interesting to note the behaviour to the cosmological constant in this theory. In the inflationary scenario, we have a positive value
in such a way that the universe has a bigger growth (but  next change
to negative one, where the universe have a small growth).

\bigskip 
\section{Quantum regimen: Wheeler-DeWitt equation}
For quantum regime, we calculate the Wheeler-DeWitt equation
\begin{equation*}\rm 
\hat{H}\Psi =\left\{ \hat{\Pi}_{\Omega }^{2}-\hat{\Pi}_{\beta _{+}}^{2}-\hat{%
\Pi}_{\beta _{-}}^{2}+\frac{6\hat{\Pi}_{\phi }^{2}\phi ^{-m}}{\omega }%
-384\pi G\mu_\gamma e^{3\Omega \left( 1-\gamma \right)} +48\Lambda e^{ 6\Omega}
\right\} \Psi =0
\end{equation*}
where the momenta operators  $\hat{\Pi}_{q}=-i\hslash \frac{\partial }{\partial q}$, $\Psi$ is the wave function of the universe,
also we choose $\hslash =1$, thus
\begin{equation*}
\hat{H}\Psi =\left\{ \left( -\frac{\partial ^{2}}{\partial \Omega^2}+Q\frac{\partial }{\partial \Omega }\right) 
+\frac{\partial ^{2}}{\partial \beta _+^2}+\frac{\partial ^{2}}{\partial \beta _-^2}
-\frac{6\phi ^{-m}}{\omega }\frac{\partial ^{2}}{\partial \phi^2 }-384\pi G\mu_\gamma e^{3\Omega \left( 1-\gamma\right)}
 +48\Lambda e^{ 6\Omega} \right\} \Psi=0, 
\end{equation*}
where we have used $\left( -\frac{\partial ^{2}}{\partial \Omega^2 }+Q\frac{\partial }{\partial \Omega }\right) $ 
for solving the factor ordering problem. Applying the separation method, using for the wave function
\begin{equation*}
\Psi ={\cal A}\left( \Omega \right) {\cal B}\left( \beta _{+}\right) {\cal C}\left( \beta
_{-}\right) {\cal D}\left( \phi \right) 
\end{equation*}
we obtain
\begin{equation}
\left\{ -\frac{1}{{\cal A}}\frac{d ^{2}{\cal A}}{d \Omega^2 }+Q\frac{1}{{\cal A}}\frac{d {\cal A}}{d \Omega }
+\frac{1}{{\cal B}}\frac{d^2{\cal B}}{d \beta _+^2}
+\frac{1}{{\cal C}}\frac{d^2{\cal C}}{d\beta_-^2}-\frac{1}{{\cal D}}\frac{6\phi ^{-m}}{\omega }\frac{d^2{\cal D}}{d \phi^2 }
-384\pi G\mu_\gamma e^{3\Omega \left( 1-\gamma \right)} +48\Lambda e^{6\Omega} \right\} =0, \nonumber
\end{equation}
yielding the following set of differential equations
\begin{eqnarray}
\rm -\frac{1}{{\cal A}}\frac{d ^{2}{\cal A}}{d \Omega^2 }+Q\frac{1}{{\cal A}}\frac{d {\cal A}}{d \Omega }
-384\pi G\mu_\gamma e^{3\Omega \left( 1-\gamma \right)} +48\Lambda e^{6\Omega}&=& \rm -a_1^2,\label{omega1}\\
\rm \frac{1}{{\cal B}}\frac{d^2{\cal B}}{d \beta _+^2}&=&\rm a_2^2, \label{beta+}\\
\rm \frac{1}{{\cal C}}\frac{d^2{\cal C}}{d\beta_-^2}&=&\rm a_3^2,\label{beta-}\\
\rm \frac{1}{{\cal D}}\frac{6\phi ^{-m}}{\omega }\frac{d^2{\cal D}}{d \phi^2 }&=&\rm a_4^2, \label{phi}
\end{eqnarray}
where $\rm a_4^2=-a_1^2+a_2^2+a_3^2$, with  $a_i^2$ separation constants. The choose of sign for these constants is 
arbitrary, in absence to initial conditions of our universe, studied under this cosmological model.

The solution for the equations  (\ref{beta+}, \ref{beta-}) have the generic form 
\begin{equation}
\rm {\cal B}=e^{\pm a_2 \beta_+}, \qquad \qquad {\cal C}=e^{\pm a_3 \beta_-},
\end{equation}

The solution for the equation (\ref{phi}) is more complicated, because depend to the constant m, which is a parameter to the S\'aez-Ballester 
theory. This equation is rewritten  as $\rm \frac{d^2{\cal D}}{d \phi^2 }- \frac{\omega a_4^2 \phi^m}{6} {\cal D}=0$, which is analog to
the equation find in  the reference \cite{andrei}, $y^{\prime\prime}-ax^n y=0$, with $a=\frac{\omega a_4^2 }{6}$ and $n=m$.
\begin{enumerate}
\item{} case m=-2 correspond to the Euler equation, who solution have the following structure \cite{andrei}

\begin{equation}
{\cal D}=\sqrt{\phi} \left\{
\begin{tabular}{lr}
$\rm  c_1\, \phi^\mu + c_2 \phi^{-\mu} $ & $\qquad$ si $\quad a>-\frac{1}{4}$ \cr
$\rm  c_1 \, + c_2 Ln \phi $ & \qquad si \quad $a=-\frac{1}{4}$ \cr
$\rm c_1\, sin\left(\mu Ln \phi \right) + c_2\, cos\left(\mu Ln \phi \right)$& \qquad si \quad $a<-\frac{1}{4}$ \cr
\end{tabular}
\right. \label{pphi2}
\end{equation}
where $\mu=\frac{1}{2} \sqrt{|1+4a|}>0$
\item{} In the case m=-4, we introduce the following transformation $z=\frac{1}{\phi}$ and $\rm \frac{u}{z}={\cal D}$, yielding to
differential equation more easy to solve
$\rm \frac{d^2 u}{dz^2}-a u=0$; so the solution become as
\begin{equation}
\rm {\cal D}=\phi \left\{
\begin{tabular}{lr}
$\rm c_1 \,sinh\left(\sqrt{a}\phi \right) +c_2\, cosh\left(\sqrt{a}\phi \right) $& \qquad si\,\, $a>0$ \cr
$\rm c_1 \,sin\left(\sqrt{|a|}\phi \right) +c_2\, cos\left(\sqrt{|a|}\phi \right) $ &\qquad si\,\, $a<0$ \cr
\end{tabular}
\right. \label{pphi4}
\end{equation} 
the case a=0, is descarted, because this imply that  $\omega=0$, yielding to the Einstein theory.

\item{} When the m parameter satisfy the relation $\frac{2}{m+2}=2n+1$, where $n$ is an integer, the general solution take the form
\begin{equation}
{\cal D}=\rm  \phi\left\{
\begin{array}{ll}
\rm\left(\phi^{1-2q}\frac{d}{d\phi}\right)^{n+1}
\left[D_6 Exp\left( \sqrt{\frac{\omega}{6}}\frac{\phi^q}{q} \right)+D_7
Exp\left(-\sqrt{\frac{\omega}{6}}\frac{\phi^q}{q}\right)\right] &\rm \qquad si\quad n\geq0 \\
\rm \left(\phi^{1-2q}\frac{d}{d\phi}\right)^{-n}\left[D_6 Exp\left(\sqrt{\frac{\omega}{6}}\frac{\phi^q}{q}\right)
+D_7 Exp\left(-\sqrt{\frac{\omega}{6}}\frac{\phi^q}{q}\right)\right]\quad & \rm \qquad si\quad n<0
\end{array}
\right.
\end{equation}
where  $D_6, D_7$ are integration constants and $q=\frac{m+2}{2}=\frac{1}{2n+1}$.

\item{} General solution for any m,  the solution is expressed in terms of the Bessel function and modified Bessel function, 
for the field $\phi$

\begin{equation}
\rm {\cal D}= \sqrt{\phi}\, Z_\nu \left(\frac{\sqrt{a}}{q} \phi^q \right),
\end{equation}
where $\rm Z_\nu$ is a generic Bessel function, $\nu=\frac{1}{2q}$ is  the order to the  corresponding Bessel function, 
$\rm q=\frac{1}{2}\left(m+2 \right)$.
 If $a<0$ imply that  $\omega<0$, $\rm Z_\nu$ become the modified Bessel function, $(I_\nu, K_\nu)$. When $a>0, \to w>0$, 
 $\rm Z_\nu \to (J_\nu,Y_\nu)$.
\end{enumerate}

On the other hand, the equation (\ref{omega1}) does have not general solution, then this is solved for particular cases in the $\gamma$ parameter,
and for this, is rewritten in the following form
\begin{equation}
\rm \frac{d ^{2}{\cal A}}{d \Omega^2 }-Q\frac{d {\cal A}}{d \Omega }
+\left(384\pi G\mu_\gamma e^{3\Omega \left( 1-\gamma \right)} -48\Lambda e^{6\Omega}- a_1^2\right){\cal A}=0,
\end{equation}
\begin{enumerate}
\item{} Any factor ordering Q and the inflation phenomenom $\gamma=-1$
\begin{equation}
\rm \frac{d ^{2}{\cal A}}{d \Omega^2 }-Q\frac{d {\cal A}}{d \Omega }+\left[b_{-1} e^{6\Omega}- a_1^2\right]{\cal A}=0, \qquad b_{-1}=384\pi G\mu_{-1}  -48\Lambda
\end{equation}
making the transformations $z=\frac{\sqrt{b_1}}{3}e^{3\Omega}$ and ${\cal A}=z^{\frac{Q}{6}} \Phi(z)$ we arrive at Bessel differential
equation for the function $\Phi$. With this
the general solution become  \cite{andrei}
\begin{equation}
\rm {\cal A}= \left(\frac{\sqrt{b_{-1}}}{3}e^{3\Omega} \right)^{\frac{Q}{6}}
Z_\nu \left( \frac{\sqrt{b_{-1}}}{3}e^{3\Omega} \right), \qquad \nu=\pm \frac{1}{6}\sqrt{Q^2+4a_1^2}
\end{equation}
where $\rm Z_\nu$ is a generic Bessel function. If $\rm b_{-1}>0$, we have the ordinary Bessel function, in other case,
will be the modified  Bessel function.
\item{} factor ordering Q=0 and $\gamma=0$
\begin{equation}
\rm \frac{d ^{2}{\cal A}}{d \Omega^2 }-\left( 48\Lambda e^{6\Omega} - b_0 e^{3\Omega}+ a_1^2\right){\cal A}=0,
\end{equation}
making the transformation $\rm z=e^{3\Omega}$ and $R=z^{-\frac{a_1}{3}} {\cal A}$, is carried to
equation
$\rm 9z\frac{d^2 R}{dz^2}+9\left(\frac{2}{3}a_1+1 \right)\frac{dR}{dz}-\left(48\Lambda z-b_0  \right) R=0$
who solution is constructed by the degenerate hypergeometric function $\rm F_1(a,b;z)$ \cite{andrei}
\begin{equation}
\rm {\cal A}=e^{a_1 \Omega} Exp\left[\frac{2}{3}\sqrt{6\Lambda}e^{a_1\Omega} \right]\,
F_1\left(\frac{B(k)}{18k},\frac{2}{3}a_1+1;\frac{e^{3\Omega}}{\lambda}\right),
\end{equation}
where $\lambda=-\frac{1}{2k}$, $\rm k=\frac{2}{3}\sqrt{6\Lambda}$, $\rm B(k)=9k(\frac{2}{3}a_1+1)+b_0$.

\item{} Any factor ordering  and stiff matter $\gamma=1$
\begin{equation}
\rm \frac{d ^{2}{\cal A}}{d \Omega^2 }-Q\frac{d {\cal A}}{d \Omega }
+\left[b_1-48\Lambda e^{6\Omega}\right]{\cal A}=0, \qquad b_1=384\pi G\mu_1  -a_1^2
\end{equation}
in similar way that the first case, the transformations $z=4\sqrt{-\frac{\Lambda}{3}}\, e^{3\Omega}$ and
${\cal A}=z^{\frac{Q}{6}}\Phi(z)$, the differential Bessel function appear for the function $\Phi$, then we have the general solution 
\begin{equation}
{\cal A}= \left(4\sqrt{-\frac{\Lambda}{3}}\, e^{3\Omega} \right)^{\frac{Q}{6}}
Z_\nu\left(4\sqrt{-\frac{\Lambda}{3}}\, e^{3\Omega} \right), \qquad \nu=\pm \frac{1}{6}\sqrt{Q^2-4b_1}.
\end{equation}
with  $\Lambda<0$, having the ordinary Bessel function; In the case when the 
factor ordering become Q=0, we have the same Bessel function, but the
cosmological constant could be positive or negative, yielding to modified Bessel function and ordinary Bessel function, 
respectively, and the order become imaginary in both cases $\nu=\pm i\frac{\sqrt{b_1}}{3}$.

\end{enumerate}
\section{conclusions}
One equivalent density lagrangian was build in order to apply the quantum regime in the S\'aez-Ballester theory, in where
the constant $\omega$ can be used to have a lorenzian (-1,1,1,1) or seudo-lorenzian (-1,-1,1,1) signature when we build the
Wheeler-DeWitt equation. The values for this parameter in the classical one is dictated when we apply the condition that
we must have real functions, which is encoded in the parameter $a$, equations (\ref{pphi2},\ref{pphi4}).
In this sense, the classical and quantum exact solutions were found for the cosmological Bianchi type I model in the frame 
of S\'aez-Ballester theory
for the scenarius in the $\gamma$ parameter $\{-1,0,1\}$. 
The presense of the exotic field $\phi$ does not delay the anisotropic behaviour in this model. Moreless, the classical behaviour of this field
for large value in the parameter $\omega$, it is similar to the anisotropic parameters $\beta_\pm$.

\acknowledgments{ \noindent This work was partially supported by CONACYT 
grants 51306 and 62253. DINPO (2009-2010) and PROMEP grants UGTO-CA-3 and UGTO-PTC-085. 
This work is part of the collaboration within the Instituto Avanzado de
Cosmolog\'{\i}a. Many calculations where done by Symbolic Program REDUCE 3.8.}

\end{document}